\documentclass{epl}
\usepackage{graphicx,amsmath,amssymb}
\usepackage{bm}

\title{Observation of macroscopic Landau--Zener transitions
in a superconducting device}
\shorttitle{Observation of macroscopic LZ transitions}
\author{A. Izmalkov\inst{1,2} \and M.~Grajcar\inst{1,3} \and E.~Il'ichev\inst{1}\thanks{E-mail: \email{ilichev@ipht-jena.de}} \and N.~Oukhanski\inst{1} \and Th.~Wagner\inst{1} \and H.-G.~Meyer\inst{1} \and W.~Krech\inst{4} \and M.~H.~S.~Amin\inst{5} \and Alec Maassen van den Brink\inst{5} \and A.~M.~Zagoskin\inst{5,6}}
\institute{
  \inst{1} Institute for Physical High Technology, P.O. Box 100239, D-07702 Jena, Germany\\
  \inst{2} Moscow Engineering Physics Institute (State University), Kashirskoe shosse 31,\\ 115409 Moscow, Russia\\
  \inst{3} Department of Solid State Physics, Comenius University, SK-84248 Bratislava,\\ Slovakia\\
  \inst{4} Friedrich Schiller University, Institute of Solid State Physics, D-07743 Jena,\\ Germany\\
  \inst{5} D-Wave Systems Inc., 320-1985 West Broadway, Vancouver, B.C., V6J 4Y3 Canada\\
  \inst{6} Physics and Astronomy Dept., The University of British Columbia,\\ 6224 Agricultural Rd., Vancouver, B.C., V6T 1Z1 Canada
}
\shortauthor{A. Izmalkov\etal}

\begin{document}

\maketitle

\begin{abstract}
A two-level system traversing a level anticrossing has a small probability to make a so-called Landau--Zener (LZ) transition between its energy bands, in deviation from simple adiabatic evolution. This effect takes on renewed relevance due to the observation of quantum coherence in superconducting qubits (macroscopic ``Schr\"odinger cat" devices). We report an observation of LZ transitions in an Al three-junction qubit coupled to a Nb resonant tank circuit.
\end{abstract}

In analogy to their classical counterparts, qubits are effectively two-level systems, with a time-dependent bias enabling one-qubit gate operations. Besides their computational use, this makes them suitable for studying Landau--Zener (LZ) transitions\cite{Garanin,Feigelman} (see below eq.~(\ref{eq01})). One prominent qubit is a superconducting loop with low inductance~$L$, interrupted by three Josephson junctions (a 3JJ qubit)\cite{Mooij99}. Its Josephson energy, $U_\mathrm{J}=\sum_{j=1}^{3}E_{\mathrm{J}j}(\phi_j)$, depends on the phase differences~$\phi_j$ across the junctions. In a small loop, due to magnetic flux quantization, only two $\phi_j$'s are independent.

\begin{figure}
\centerline{\includegraphics[width=3in]{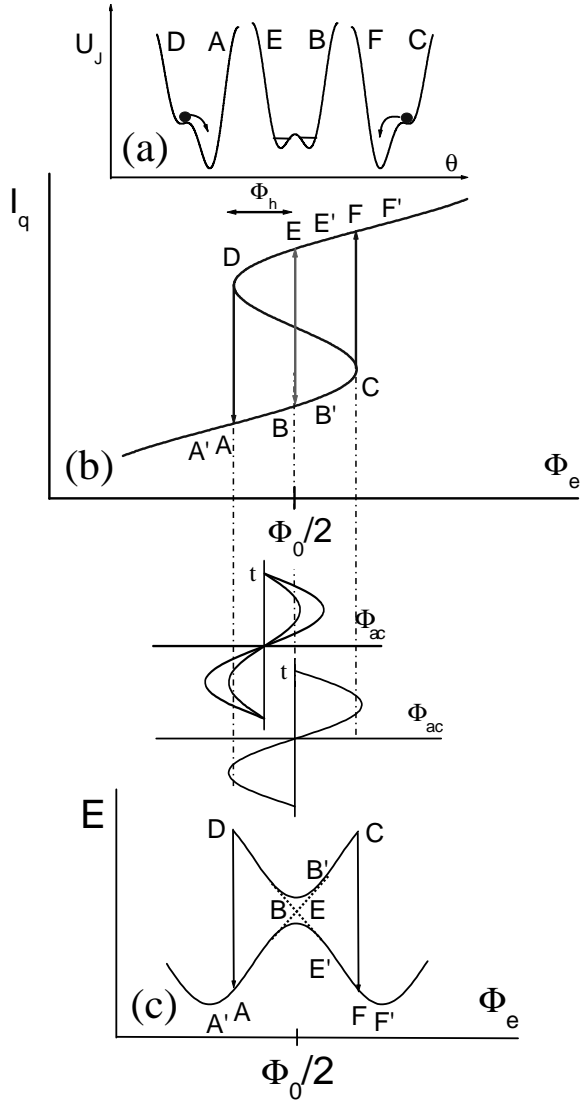}}
\caption{(a) Josephson energy vs phase $\theta$ along the saddle-point
trajectory, for three values of the external qubit flux~$\Phi_\mathrm{e}$. (b)~Persistent current $I_\mathrm{q}$ vs~$\Phi_\mathrm{e}$. In the ``classical'' case (bottom sine), the hysteresis is centred around the degeneracy point $\Phi_\mathrm{e}=\frac{1}{2}\Phi_0$; in the ``quantum'' one (top sine), only one hysteretic jump has to be enclosed together with the anticrossing, so half the bias amplitude suffices. (c)~Quantum energy levels of the qubit vs external flux. The dashed lines correspond to the classical potential minima. In all graphs, the states $A$, $B$, $C$ correspond to, say, $\psi_\mathrm{L}$ with left-rotating supercurrent. As $\Phi_\mathrm{e}$ is increased, these lose classical stability in favour of the corresponding states~$\psi_\mathrm{R}$, denoted by $D$, $E$, and~$F$.} \label{fig1}
\end{figure}

The two minima in $U_\mathrm{J}(\phi_1,\phi_2)$ correspond to the
qubit states $\psi_\mathrm{L}$ and $\psi_\mathrm{R}$, carrying opposite
supercurrents around the loop. These become degenerate in the
presence of an external magnetic flux
$\Phi_\mathrm{e}=\frac{1}{2}\Phi_0$ ($\Phi_0\equiv h/2e$~is the flux
quantum). The potential $U_\mathrm{J}$ is sketched in fig.~\ref{fig1}a.

The Coulomb energy $U_Q$ of charges in the junctions
introduces quantum uncertainty in the~$\phi_j$. Hence, near
degeneracy the system can tunnel between the two potential minima.
(Since $U_Q \ll U_\mathrm{J}$, we deal with a so-called
\emph{flux} qubit; $U_Q \gg U_\mathrm{J}$ yields a \emph{charge}
qubit. Coherent tunneling was demonstrated in both.)

In the basis $\{\psi_\mathrm{L},\psi_\mathrm{R}\}$ and near $\Phi_\mathrm{e}=\frac{1}{2}\Phi_0$, the qubit can be described by the Hamiltonian
\begin{equation}\label{eq01}
  H(t)= - \frac{1}{2}[W(t)\sigma_z +
  \Delta\sigma_x]\,;
\end{equation}
$\sigma_x$, $\sigma_z$ are Pauli matrices and $\Delta$ is the tunneling splitting. LZ originally considered a linear bias sweep $W(t)=\lambda t$\cite{LZ}. Then, the instantaneous eigenstates $\psi_\pm(t)$ have energies $\pm\frac{1}{2}\Omega(t)\equiv\pm\frac{1}{2}\sqrt{\Delta^2{+}W(t)^2}\approx\pm\frac{1}{2}\lambda|t|$ as $|t|\to\infty$, and the LZ transition probability from the ground state $\psi_-(t{=}{-}\infty)$ to the excited state $\psi_+(t{=}\infty)$ can be obtained exactly, as $P_\mathrm{LZ}=\exp(-\pi\Delta^2\!/2\hbar\lambda)$.

We study LZ events for a periodic $W(t)$ through the associated energy loss. Indeed, classically the system of fig.~\ref{fig1} is hysteretic, and in the presence of a flux $\Phi_\mathrm{e}(t) = \Phi_\mathrm{dc}-\nobreak\Phi_\mathrm{ac}\cos(2\pi\nu t)$ with amplitude $\Phi_\mathrm{ac}>\Phi_\mathrm{h}$, the half-width of the hysteresis loop $ACFD$ (fig.~\ref{fig1}b)\cite{fc}, it will register losses proportional to the loop area in every period~$\nu^{-1}$, as long as
\begin{equation}
  |\Phi_\mathrm{dc}-{\textstyle\frac{1}{2}}\Phi_0|<
  \Phi_\mathrm{ac}-\Phi_\mathrm{h}\;.\label{plateau}
\end{equation}
These losses occur at the jumps from $\psi_+$ to $\psi_-$ at the ends of the loop.

For $|\Phi_\mathrm{dc}-\frac{1}{2}\Phi_0|>\Phi_\mathrm{ac}-\Phi_\mathrm{h}$, the classical hysteresis loop will not close. Losses are nevertheless possible due to tunneling: a nonzero area in the configuration plane can be enclosed, \emph{e.g.}\ along $A'BEE'DAA'$ in fig.~\ref{fig1}b for $\Phi_\mathrm{dc}<\frac{1}{2}\Phi_0$, through consecutively flux tunneling ($BE$), a LZ transition (at $E$ along $E'D$), and a classical jump ($DA$). The LZ event prevents the system from returning along $E'EBA$ by taking it out of its (quantum) ground state, keeping it in the classical metastable state (dashed lines in fig.~\ref{fig1}c). The order of tunneling and LZ events can also be reversed, as in the companion process $A'B'BEDAA'$; in our experiment, interference between the two paths cannot be resolved. Analogous processes, involving the other half $FEBC$ of the hysteresis loop instead of $ABED$, take place for $\Phi_\mathrm{dc}>\frac{1}{2}\Phi_0$, so that these ``quantum'' hysteresis loops can close if $\Phi_\mathrm{h}-\Phi_\mathrm{ac}<|\Phi_\mathrm{dc}-\frac{1}{2}\Phi_0|<\Phi_\mathrm{ac}$. Finally, for still smaller $\Phi_\mathrm{ac}<\frac{1}{2}\Phi_\mathrm{h}$, loss through this mechanism is impossible.

Since after the cycle $A'BEE'DAA'$ the system has returned to its initial state $\psi_-$, the losses are proportional to the probability of quantum evolution from $\psi_-$ to $\psi_+$ before the classical jump along $DA$. Therefore, measuring losses indeed probes the LZ effect (generalized to nonlinear sweep and finite transition time).

There is a major difference between the ``classical" and ``quantum" hysteresis loops above. For the former, the loss occurs whenever the sweep cycle covers the loop, as determined by eq.~(\ref{plateau}). For the latter, either two consecutive tunneling or two LZ events would bring the system back along $BA'$ in the above example. With the two mentioned contributing paths, the net dissipation is therefore proportional to
\begin{equation}
  P_\mathrm{loss}=2P_\mathrm{LZ}(1-P_\mathrm{LZ})\;,\label{Ploss}
\end{equation}
and vanishes if $P_\mathrm{LZ}$ is either too small or too large. Due to our system's moderately small $\Delta$ and the exponential dependence of $P_\mathrm{LZ}$ on the sweep rate (see below eq.~(\ref{eq01})), in practice this makes the ``quantum" losses observable only if the bias sweep narrowly overshoots the anticrossing, \emph{i.e.}, if
\begin{equation}
  0<\Phi_\mathrm{ac}-|\Phi_\mathrm{dc}{-}{\textstyle\frac{1}{2}}\Phi_0|\ll\Phi_0\;,
  \label{peak}
\end{equation}
when $\dot{\Phi}_\mathrm{e}|_{\Phi_0/2}$ is small. When plotting $P_\mathrm{loss}(\Phi_\mathrm{dc})$ for, say, $\Phi_\mathrm{ac}>\Phi_\mathrm{h}$, one therefore expects a ``classical" \emph{plateau} flanked by two ``quantum" \emph{peaks}, at a distance $\Phi_\mathrm{h}$ from the former's edges. The latter present an interesting analogy to the atomic collisions in terms of which LZ transitions were originally discussed\cite{LZ}. Namely, such a collision is inelastic if the system jumps exactly once near an anticrossing of electronic levels (in a given angular-momentum sector), which it traverses both during approach and recoil of the two nuclei. See fig.~31 in ref.\cite{LL3}, and compare its eq.~(90,15) to eq.~(\ref{Ploss}) above.

For an experimental realization, we have used a technique similar to rf-SQUID readout\cite{Silver67,Ilichev01}. Namely, the effective inductance of a superconducting loop with Josephson junctions depends on the flux threading the loop. The loop is inductively coupled to a parallel resonant tank circuit\cite{Silver67}. The tank is fed a monochromatic rf signal, close to its resonant frequency $\nu$. Then both (rms) magnitude $V_\mathrm{T}$ and phase of the tank voltage will strongly depend on (A)~the shift in resonant frequency due to the change of the loop inductance by the external flux\cite{state}, and (B)~losses in the loop caused by field-induced transitions between the two quantum states. Thus, the tank both applies the probing field to the qubit, and detects its response.

The output signal depends on the tank's quality factor $Q$. Using superconducting coil, values as high as $Q\sim 10^3$--$10^4$ can be obtained, leading to high readout sensitivity, \emph{e.g.}, in rf-SQUID magnetometers\cite{Danilov}. Such a tank can therefore be used to probe flux qubits\cite{Greenberg02a,Greenberg02b}. Even a weak coupling to a qubit will substantially decrease the effective $Q$ if the qubit is in the dissipative regime, leading to a dip in $V_\mathrm{T}$ as mentioned under (B) above.

For the tank, we prepared square-shaped Nb pancake coils on oxidized Si substrates. Predefined alignment marks allow placing a qubit in the centre of the coil. For flexibility, only the coil was made lithographically. We use an external capacitance $C_\mathrm{T}$ to be able to change~$\nu$. The line width of the 30 coil windings was 2~$\mu$m, with a 2~$\mu$m spacing. The resonant properties of the tank ($L_\mathrm{T}\approx138$~nH, $C_\mathrm{T}\approx470$~pF) used for the reported measurements were obtained from the voltage--frequency characteristic as $\nu=19.771$~MHz and $Q\approx1680$.

\begin{figure}[t]
\centerline{\includegraphics[width=8cm]{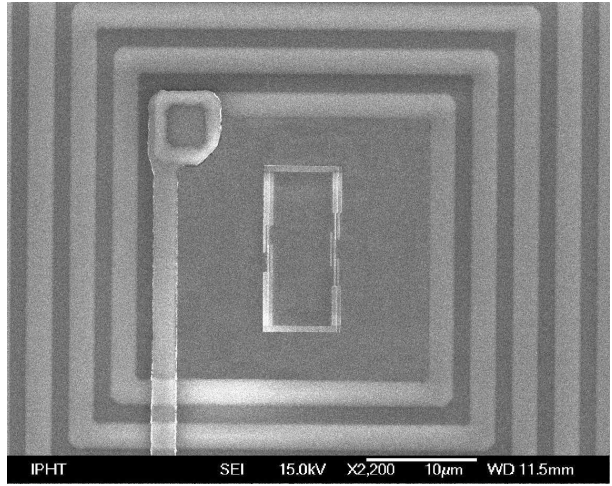}}
\caption{Electron micrograph of the qubit at the centre of the tank coil.}\label{sample}
\end{figure}

The 3JJ qubit structure was manufactured out of Al in the middle of the coil by conventional shadow evaporation; see fig.~\ref{sample}. The junctions have areas $\approx130\times620$, $120\times600$, and $110\times610$~nm$^2$ respectively. From the known properties of the fabrication process, the critical current and capacitance of the largest junction are then $\approx240$~nA and $\approx4$~fF respectively, and proportional to the area for the other ones. The loop area was 90~$\mu\mathrm{m}^2$, with $L=39$~pH. The sample was placed at the mixing chamber of a dilution refrigerator with a base temperature of $\sim10$~mK

We measured $V_\mathrm{T}$ by a three-stage cryogenic amplifier, placed at $\approx2$~K and based on commercial pseudomorphic high electron mobility transistors. It was slightly modified from the version in ref.\cite{Oukhansky02} in order to decrease its
back-action on the qubit. The input-voltage noise was $<0.6$~nV\!/$\sqrt{\mathrm{Hz}}$ in the range \mbox{1--25}~MHz. The noise temperature was $\sim 200$~mK at $\approx20$~MHz. The effective qubit temperature (depending on the amplifier's back-action, the cooling power of the fridge, etc.) should be considerably lower because of the small tank--qubit coupling $M/\sqrt{LL_\mathrm{T}}\approx1.3\cdot10^{-2}$ ($M$ is the mutual inductance), as was verified experimentally in ref.~\cite{state} for a comparable sample.

\begin{figure}[t]
\centerline{\includegraphics[width=8cm]{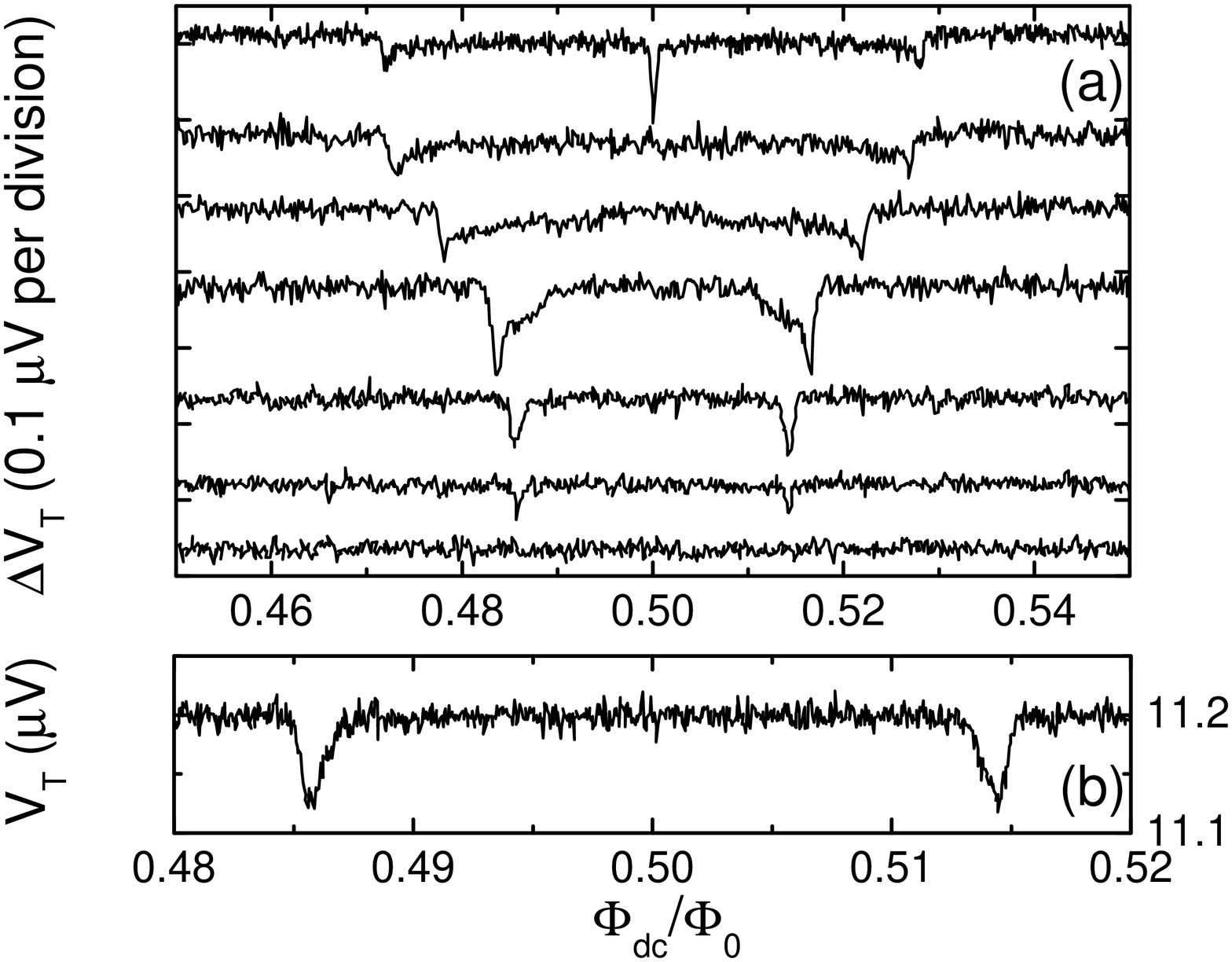}}
\vspace{3mm}
\centerline{\includegraphics[width=8cm]{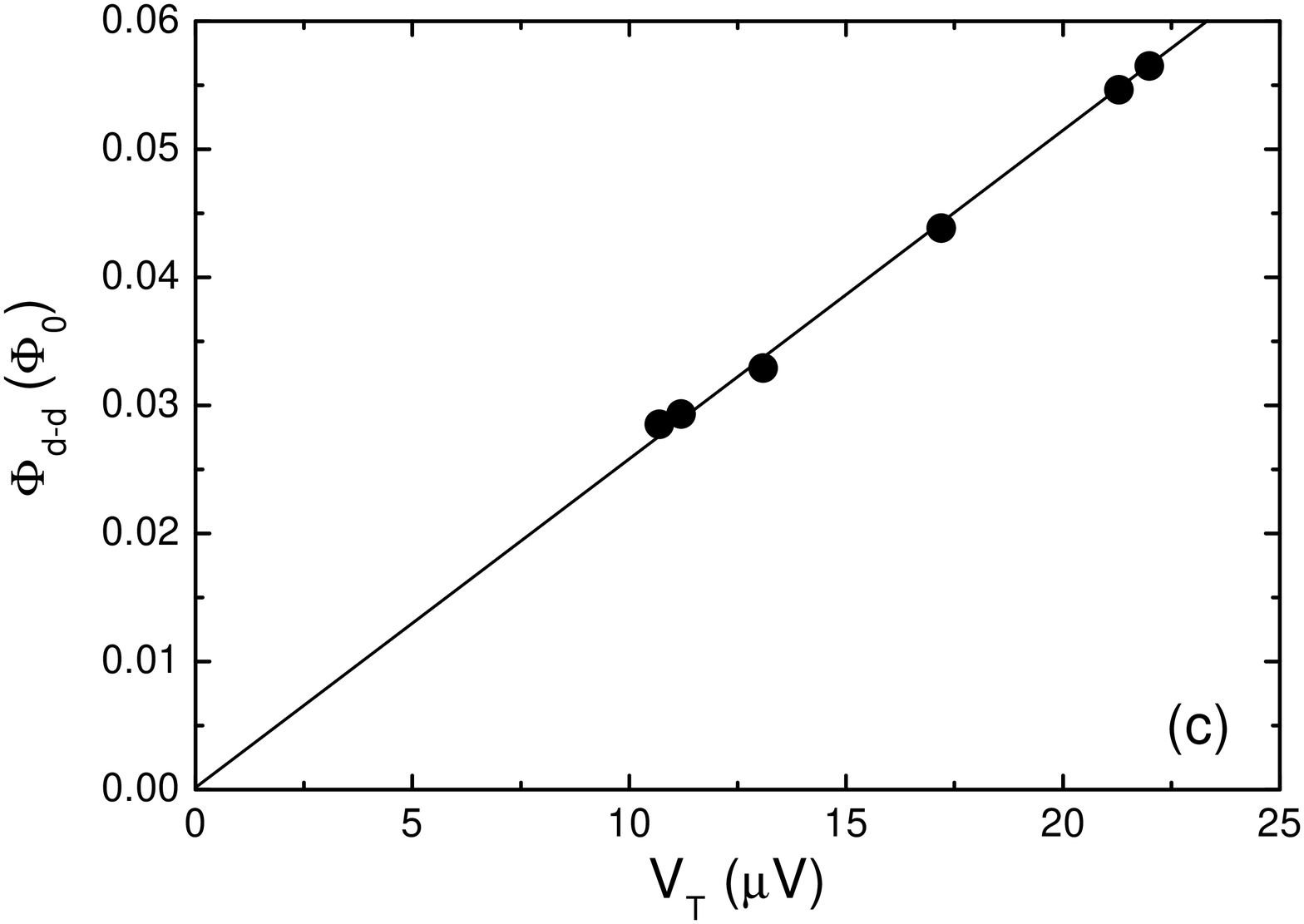}}
\caption{Tank voltage vs flux bias near the degeneracy point~$\frac{1}{2}\Phi_0$. (a)~From the lower to the upper curve, the driving voltage is 10.2, 10.7, 11.2, 13.1, 17.2, 21.3, 22.0~$\mu$V rms (data vertically shifted for clarity). (b)~Close-up of the 11.2~$\mu$V curve. (c)~The distance between the two dips $\Phi_\text{d--d}$ for the upper six curves in (a) vs bias amplitude, in agreement with $\Phi_\text{d--d}=2\Phi_\mathrm{ac}$ as predicted by eq.~(\ref{peak}).}\label{fig2}
\end{figure}

Reproducibility under, \emph{e.g.}, thermal cycling was excellent throughout. Results for small driving voltage are shown in fig.~\ref{fig2}. For the smallest voltages no dissipative response is observed; the two ``quantum'' peaks appear around $10.7~\mu$V\cite{Fac}, and subsequently move apart without broadening significantly, with a separation proportional to the ac bias amplitude. The ``classical'' peak appears in the centre, and with an ac bias threshold \emph{double} the one of the quantum peaks---both as predicted above.

\begin{figure}
\centerline{\includegraphics[width=8cm]{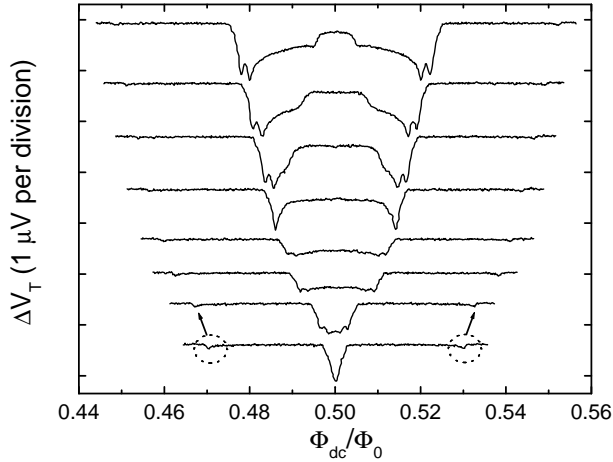}}
\caption{Same as in fig.~\ref{fig2}, for voltages 23.8, 25.9, 30.0, 32.1, 34.4, 36.4, 38.5, 41.0~$\mu$V rms from bottom to top. With increasing tank-voltage amplitude, the small dips (marked for $23.8~\mu$V) move away from the degeneracy point.}\label{fig3}
\end{figure}

The development of these structures can be readily followed in the lowest four traces in fig.~\ref{fig3}. On the scale of the figure, the quantum peaks are tiny but clearly visible; they move apart at a fixed distance from the widening classical plateau.

Finally, for the remaining traces in fig.~\ref{fig3}, the bias amplitude is large enough to cause substantial deviations from our simple two-state model. These repeatable results are given here for reference. Detailed modeling can presumably be based on a master equation\cite{opa}, in which the (time-dependent) states follow from the qubit's band structure\cite{Mooij99}; the transition rates will account for both dissipative and LZ processes, where the latter contribution is localized at the level anticrossings. While this large-amplitude regime is irrelevant for qubit operation, some of its features, \emph{e.g.}\ the distinct, robust two-pronged structures at the predicted edge of the classical plateau, are intriguing and warrant further investigation.

In conclusion, we have observed Landau--Zener transitions in a macroscopic superconducting system: an Al flux qubit coupled to a Nb resonant tank. The latter played dual, control and readout, roles. The impedance readout technique allows detecting a quantum process, using a \emph{low-frequency}, \emph{dissipative} method.

\acknowledgments
We thank \textsc{A.~J. Leggett} (who also commented on the manuscript), \textsc{A.~N. Omelyanchouk}, \textsc{D.~E. Sheehy} and \textsc{A.~Yu.\ Smirnov} for fruitful discussions.


\begin{thebibliography}{99}
\bibitem{Fac} According to our model, this driving voltage should correspond to $\Phi_\mathrm{ac}=|\Phi_\mathrm{dc}{-}\frac{1}{2}\Phi_0|=\frac{1}{2}\Phi_\mathrm{h}\approx0.014\Phi_0$. While this is certainly correct up to a factor of two, and deviations $\sim10\%$ due to parasitic effects would be unsurprising, uncertainties in the calibration of the cryogenic amplifier at present preclude a stronger statement.
\end{thebibliography}
\end{document}